\documentclass[11pt, a4paper]{article}
\usepackage{jcappub}

\usepackage{color}
\usepackage{booktabs}
\usepackage{multirow}
\usepackage{amssymb}
\usepackage[export]{adjustbox}
\usepackage{floatrow}
\newfloatcommand{capbtabbox}{table}[][3.5in]
\newfloatcommand{capbfigbox}{figure}[][2.3in]
\usepackage{blindtext}
\usepackage{amsmath}
\usepackage{booktabs}
\usepackage{aas_macros}
\usepackage{epsfig}
\usepackage{graphicx}
\usepackage{subfigure}

\def\lsim{\mathrel{\raise.3ex\hbox{$<$\kern-.75em\lower1ex\hbox{$\sim$}}}}
\def\gsim{\mathrel{\raise.3ex\hbox{$>$\kern-.75em\lower1ex\hbox{$\sim$}}}}

\begin{document}

\title{Interpreting the GeV-TeV Gamma-Ray Spectra of Local Giant Molecular Clouds using GEANT4 Simulation}

\author[a,]{Abhijit Roy,}
\author[a,b]{Jagdish C. Joshi,}
\author[c,]{Martina Cardillo,}
\author[d,]{and Ritabrata Sarkar}

\affiliation[a]{Aryabhatta Research Institute of Observational Sciences (ARIES), \\ Manora peak, Beluwakhan, Uttarakhand 263002, India}

\affiliation[b]{Centre for Astro-Particle Physics (CAPP) and Department of Physics, University of Johannesburg, PO Box 524, Auckland Park 2006, South Africa}

\affiliation[c]{INAF - Istituto di Astrofisica e Planetologia Spaziali (IAPS), \\ Via del Fosso del Cavaliere 100, 00133 Roma, Italy}
\affiliation[d]{Institute of Astronomy Space and Earth Science (IASES), \\ 316, AJ Block, Sector II, Bidhannagar, Kolkata, West Bengal 700091, India}

\emailAdd{abhijitroy@aries.res.in (AR)}
\emailAdd{jagdish@aries.res.in (JCJ)}
\emailAdd{martina.cardillo@inaf.it (MC)}
\emailAdd{ritabrata.s@gmail.com (RS)}

\newcommand\mc[1]{{\bf \color{magenta} #1}}
\newcommand\rs[1]{{\bf \color{cyan} #1}}



\abstract{Recently, the Fermi-LAT gamma-ray satellite has detected six Giant Molecular Clouds (GMCs) located in the Gould Belt and the Aquila Rift regions. In half of these objects (Taurus, Orion A, Orion B), the observed gamma-ray spectrum can be explained using the Galactic diffused Cosmic Ray (CR) interactions with the gas environments. In the remaining three GMCs (Rho Oph, Aquila Rift, Cepheus), the origin of the gamma-ray spectrum is still not well established. We use the GEometry ANd Tracking (GEANT4) simulation framework in order to simulate gamma-ray emission due to CR/GMC interaction in these three objects, taking into account the gas density distribution inside the GMCs. We find that propagation of diffused Galactic CRs inside these GMCs can explain the Fermi-LAT detected gamma-ray spectra. Further, our estimated TeV-PeV fluxes are consistent with the HAWC upper limits, available for the Aquila Rift GMC. As last step, we compute the total neutrino flux estimated for these GMCs and compare it with the IceCube detection sensitivity.

}

\maketitle

\section{Introduction}
Cosmic Rays (CRs) are relativistic particles, mainly protons and ions, that fill our Galaxy. Their spectrum spans from some MeV to $10^{20}$ eV with a Galactic component contributing at least up to $\sim 3 \times 10^{15}$ eV or 3 PeV \citep{2013FrPhy...8..748G}. The transition energy where the source class changes from Galactic to extra-galactic is still debatable, and it is expected to be in between $10^{17}$ to $10^{19}$ eV \citep{2013FrPhy...8..748G}. SNRs are one of the most plausible sources of Galactic CRs, mainly because they can supply the required energy budget for Galactic CRs (GCRs) and are the perfect environment for first-order Fermi acceleration mechanisms. One of the main channels to confirm CR origin is non-thermal High Energy (HE) gamma-ray emission that, however, can be produced both by electrons, via Bremsstrahlung or Inverse Compton (leptonic processes), and by protons, via $p-p$ interactions (hadronic processes). The necessary but not sufficient condition for a source to be a CR accelerator is gamma-ray emission above 100 TeV (corresponding to a particle of PeV energies) but in all known SNRs, a cutoff or break has been observed at $E<100$ TeV \citep{funk2017high}, which makes them less optimistic candidates for PeVatrons. Binary systems in our Galaxy can inject CR particles above PeV via magnetic induction, but they do not provide a sufficient energy budget to explain the observed CR flux on Earth \citep{Guenduez18,2020AnA...635A.167H, 2020AnA...635A.144W}. As per the population of stellar-mass black holes in X-ray binaries, \cite{2023MNRAS.tmp.1839K} found that the contribution of these objects to the TeV CR spectra is 50 $\%$ and the contribution in the PeV regime is still uncertain.  Massive stars and pulsar wind nebula can be other prototype Galactic sources of CRs \citep{2019NatAs...3..561A, Rou_Wg_Crab}. Investigations of all these Galactic sources provide us with unique opportunities to study and reveal their capability as potential CR accelerators. 
One of the main problems is that gamma-ray emission above 100 TeV could be due both to electrons and protons, consequently, this is not a sufficient condition to define a source as CR accelerator. In this context, the operation of neutrino detectors becomes fundamental. Neutrino detection from Galactic sources is the key to revealing CR sources because neutrinos are produced only by hadronic processes (CR interaction with target material). Multi-messenger observations and modelling of Galactic sources are important to identify potential CR sources \citep{2010AnA...511A...8B, 2015APh6580M, 2019NCimR..42..549B, 2019IJMPD..2830022G, 2021Univ324C, 2022arXiv221210677V, 2023MNRAS.521.1144S}. So far there are no confirmed detection of neutrino events from Galactic sources \cite{IC2020PhRv124e1103A} because of the low sensitivity of current detectors but with the future IceCube-Gen2 \citep{aartsen2021icecube} the detection statistics will be enhanced.

One of the main chances to detect hadronic gamma-ray emission at $E>100$ TeV are Giant Molecular Clouds (GMCs) that provide us with a laboratory to investigate CRs due to their large gas density environments. Indeed, their very high-density value enhances the interaction probability of CRs with the target material, and their multi-messenger observations can be used to identify any nearby CR accelerator and to analyse and understand CR transport in the Galactic environments \citep{Ahar1996AnA...309..917A}. Sources like MAGIC J1837-073 \citep{banik2021EPJC478B}, SNR W28 \citep{aharonian2008discovery, 2010MNRAS.409L..35L}, and SNR G35.6–0.4 \citep{zhang2022ApJ128Z} are some of the examples where Molecular Cloud (MC) association with a CR source has been identified. Moreover, the gamma-ray emission from isolated MCs are useful for studying the nature of the diffused spectrum of CRs at various locations in the Galaxy \citep{2017A&A...606A..22N, albert2021probing}. 
 

In this work, we focus on the interaction of the diffuse component of GCRs with some local isolated GMCs (listed in Table 1) and we investigate how gas distributions affect the produced gamma-ray emission. Re-accelerated particles could also affect the spectral shape of the observed gamma-rays from the MCs \citep{2016AnA...595A..58C}, but we ignore these effects here. 
Our starting point is the recent work by \citep{bagh20} where they analysed the Fermi-LAT gamma-ray spectra detected from six nearby GMCs, it was found that the computed gamma-ray spectra in some of them cannot be interpreted using AMS-like spectrum and additional CR sources are required to interpret the excess. In their analysis, they use an analytical model assuming a constant gas density in the whole cloud. Nevertheless, they postulated that several 100 T-Tauri stars in GMC environments could inject an extra component of CRs that, interacting with the gas medium, can produce additional gamma rays. In our work, we not only consider the impact on the $\pi^{0}$ decay gamma-ray flux due to primary diffused GCRs but also include the additional contribution due to the secondary CR nuclei produced inside the GMCs. Their total contribution towards $\pi^0$ decay gamma-rays is calculated using GEometry ANd Tracking (GEANT4) simulation.
Further, we have selected a gas distribution model for the GMCs by assuming their spherical morphology.

 
The outline of this paper is the following. In Section \ref{Sec:3GMCs}, we describe the properties of the three GMCs analysed (Rho Oph, Aquila Rift and Cepheus) and their known observations. In Section \ref{Sec:GEANT} We describe the CR spectrum, gas density model of GMCs and propagation of CRs through this gas distribution and production of secondary particles using GEANT4 simulation. We further, discuss our results in Section \ref{Sec:Discussion} and conclude it in Section \ref{Conclusions}.   

\section{Gamma-ray Excess from Local GMCs}
\label{Sec:3GMCs}

In \citep{bagh20}, the authors analysed the gamma-ray spectra of six nearby GMCs detected by Fermi-LAT i.e., Aquila Rift, Taurus, Rho Oph, Orion A, Cepheus, and Orion B. Their spectrum follows a power-law in the range, 3 GeV - 1 TeV and they try to fit it using GCR interactions with the gas density available inside these GMCs. They have used a constant gas-density model and the value of the CR injection is the same used to analyse the GCR proton distribution measured by Alpha Magnetic Spectrometer (AMS-02) \citep{aguilar2015precision}. The observed gamma-ray spectral distribution from Taurus, Orion A, and Orion B GMCs agreed well with their model but it significantly differs both in the spectral index and total flux for the remaining three objects, Rho Oph, Aquila Rift, and Cepheus. In Rho Oph, Aquila Rift, and Cepheus GMCs, the gamma-ray spectra cannot be explained by pion-decay model, if the CRs injected inside GMCs are based on the detection by AMS-02 experiment \citep{bagh20}.

In this work, our primary interest is the comprehension of the origin of this gamma-ray excess reported by Fermi-LAT in these three GMCs. The main features of these objects will be discussed in the following subsections, but the relevant information is listed in Table \ref{Tab:cloudp}, where their coordinates are tabulated in the second and third columns. The GMC radius is calculated from their apparent angular area. The mass is calculated from the factor $A = M_5/D^2_{kpc}$, where $M_5 = M/10^5M_\odot$ and $M_\odot$ is the Solar Mass, computed using data from the observation of \textit{Planck} dust opacity map at 353 GHz \citep{planck2011planck}. The systematic uncertainty associated with the calculated mass ($14\%$) comes from the observations \cite{bagh20,planck2011planck}. Since the gamma-ray flux from the GMCs is directly proportional to $M/D^2$ times the incident GCR flux \citep{casanova2010molecular}, accurate measurements of the mass (M) and distance (D) of the GMCs are crucial. The value of $M/D^2$ for the Aquila Rift is higher than that of the other GMCs, as shown in Table \ref{Tab:cloudp}, and as a result, the observed gamma-rays on Earth from this GMC is larger compared to other two objects.

\begin{table}[!ht]
{\footnotesize
\begin{tabular}{lcccccccc}
\hline\hline
Name & l & b & Distance (D) & A & Ang. Area & Mass (M) & Radius & M/D$^{2}$\\
 & (deg) & (deg) & (pc) & ($10^5 M_\odot kpc^{-2}$) & ($deg^2$) & ($M_\odot$) & (pc) & ($M_\odot/pc^{2}$)\\
\hline
Rho Oph     & 354.34 & 16.82  &125   & 3.98   & 24   &  6.22$\times 10^3$   & 6.03  & 0.40 \\
Aquila Rift & 24.14  & 12.48  & 225  & 16.02  & 104  &  8.11$\times 10^4$  & 22.59 & 1.60 \\
Cepheus     & 107.94 & 15.07  & 860  & 3.73   & 29   &  2.76$\times 10^5$ & 45.60 & 0.37 \\
\hline            
\end{tabular}
}
\caption{List of the three GMCs used in this work. The parameters l, b, Distance, A, and Angular Area are taken from \citep{bagh20}, and the remaining parameters are derived.}
\label{Tab:cloudp}
\end{table}

\subsection{Rho Oph}
The Rho Oph MC is a region in the constellation \textit{Ophiucus} and is composed of gas and dust grains. It is one of the nearest star-forming regions, located at a distance $\sim 125$
pc covering an angular area of $24$ degree$^{2}$ in the sky. The total mass of the cloud is $\sim 6.22\times10^3$ $M_\odot$. This GMC is an active star-forming region and more than hundreds of young T-Tauri stars were observed inside it \citep{bontemps2001isocam}. These stars are typically pre-main-sequence stars younger than 10 million years, with strong stellar winds, powerful magnetic fields and frequent flares \citep{appenzeller1989t}. The gamma-ray emission in the cloud region has been well studied by several space- and ground-based gamma-ray detectors like COS-B \citep{1983SSRv...36...61H}, Fermi-LAT \citep{bagh20,yang2014probing}, and HAWC \citep{albert2021probing}. Owing to the observation, several authors claimed a gamma-ray excess in correspondence of the cloud \citep{1985ICRC....1..197M, bagh20, albert2021probing} and the work by \citep{bagh20} concluded that this excess should come from the several hundreds of young T-Tauri stars contained within the clouds. However, they admit that the energy released from those stars is less than that necessary to explain the detected gamma-ray flux, hence a contribution from other sources is required to explain the excess. In addition, radio 
and X-ray observations \citep{1985ICRC....1..197M} didn't find any strong source near the cloud core that could contribute to the excess, and the scenario remains still not well understood.

\subsection{Aquila Rift}
Aquila Rift is another dense MC located very close to the Galactic plane and 
spanning through the constellation of \textit{Aquila} and \textit{Serpens}. Its total mass is $\sim$ $8.11\times10^4$ $M_\odot$ and it is located at a distance of $\sim$ 225 pc  with an angular area of 104 degree$^{2}$. For this object, Fermi-LAT observations in the energy range 3 -- 1000 GeV \citep{bagh20} and HAWC upper limits in the energy range 1 -- 100 TeV are available  \citep{albert2021probing}. The observation revealed that the expected limits on the gamma-ray flux observed by the HAWC at 95\% confidence level are about ten times higher than the expected gamma-ray flux. The gamma-ray excess detected from the Aquila Rift region cannot be explained by star-forming regions located inside the cloud because most of them are excluded from the analysis template of Fermi-LAT observation \cite{bagh20}. The gamma-ray observations are available for this object from GeV to PeV range and, consequently, we extend our calculations to PeV energies for this object.

\subsection{Cepheus}
Cepheus is another MC located very close to the Galactic plane. This cloud, situated in the constellation of \textit{Cepheus} at a distance of $\sim$ 860 pc, is the largest among the three GMCs studied in this work with a radius of $\sim$ 45.6 pc. Its mass is around $\sim$ $2.76\times10^5$ $M_\odot$. The diffuse emission of gamma-rays from this region also shows a clear excess as calculated by \citep{bagh20}. Its gamma-ray spectrum shows slightly steeper flux levels compared to the other two objects discussed above \citep{bagh20}. The origin of gamma-ray excess could be due to the injection of CRs by stellar winds as expected for this source \citep{bagh20}. In the vicinity of this object, three OB-association are present \citep{kun2008handbook}. However, in this work, we have not included the injection of these CRs inside Cepheus cloud, and only diffuse GCR propagation is investigated.


\section{Interactions of Galactic Diffused CRs with GMCs using GEANT4}
\label{Sec:GEANT}

\begin{figure}[!h]
    \centering
    \includegraphics[scale=0.85]{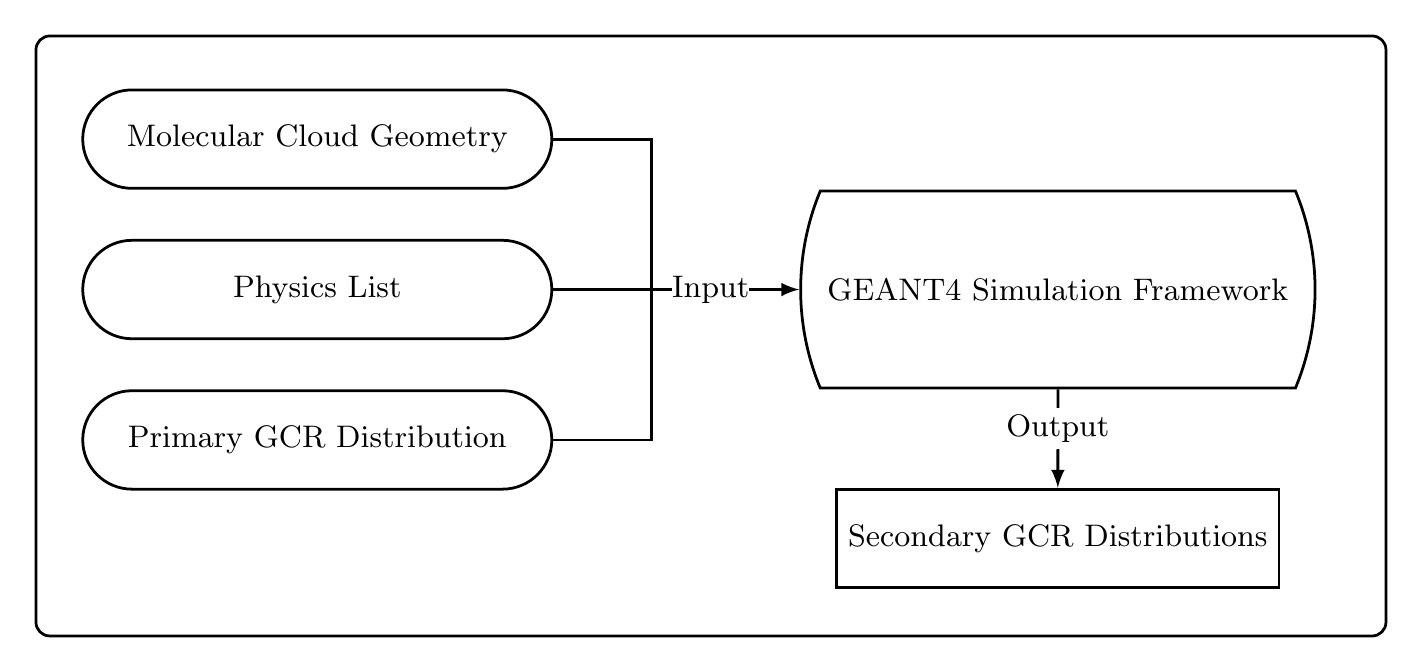}
    \caption{A flow chart diagram of the simulation. As shown here, we required as input the CR distribution in the vicinity of the GMC, the gas distribution inside the GMC, and the interaction models as defined in the physics list (for more details see Section \ref{Sec:GEANT}). Using these inputs, we estimate secondary gamma-ray and neutrino fluxes from GMCs using GEANT4. }
    \label{fig:flow_digr}
\end{figure}

GEANT4 is a Monte Carlo toolkit to simulate the interactions of particles with matter \citep{agostinelli2003geant4}. It is a very versatile tool that is widely used in many different fields of physics including nuclear physics, medical sciences, detector development, etc. Here, we have used this toolkit to simulate the gamma-ray and neutrino fluxes from local GMCs, taking into account a realistic three-dimensional mass model of the MC, physical interaction cross-sections, and primary GCR Local Interstellar Spectrum (LIS). Figure \ref{fig:flow_digr} shows a simplistic flow chart diagram of the simulation framework, and different models used in the simulation are described below.

\begin{figure}[!ht]
\centering
\subfigure{\includegraphics[width=0.34\linewidth]{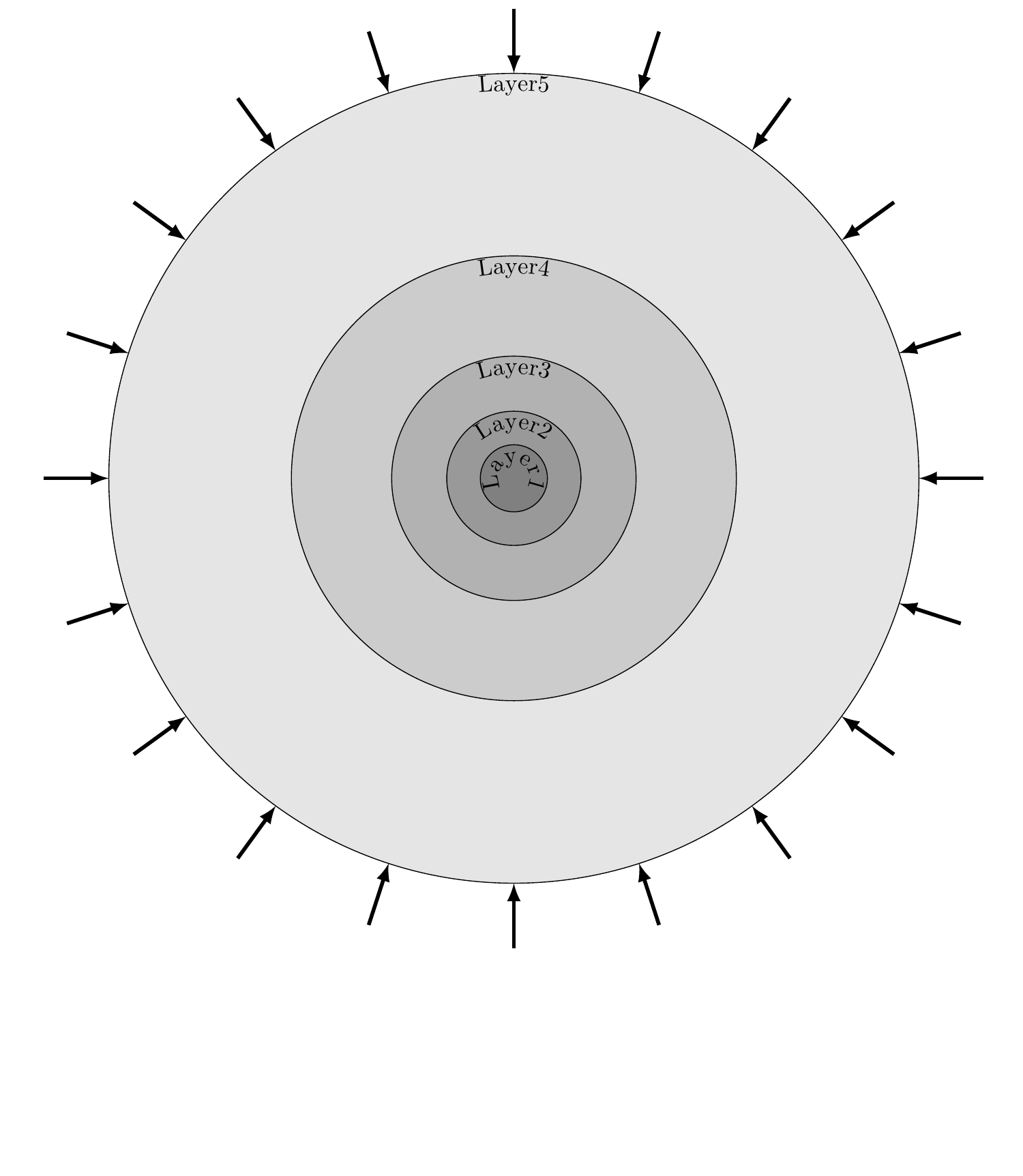}}
\subfigure{\includegraphics[width=0.56\linewidth]{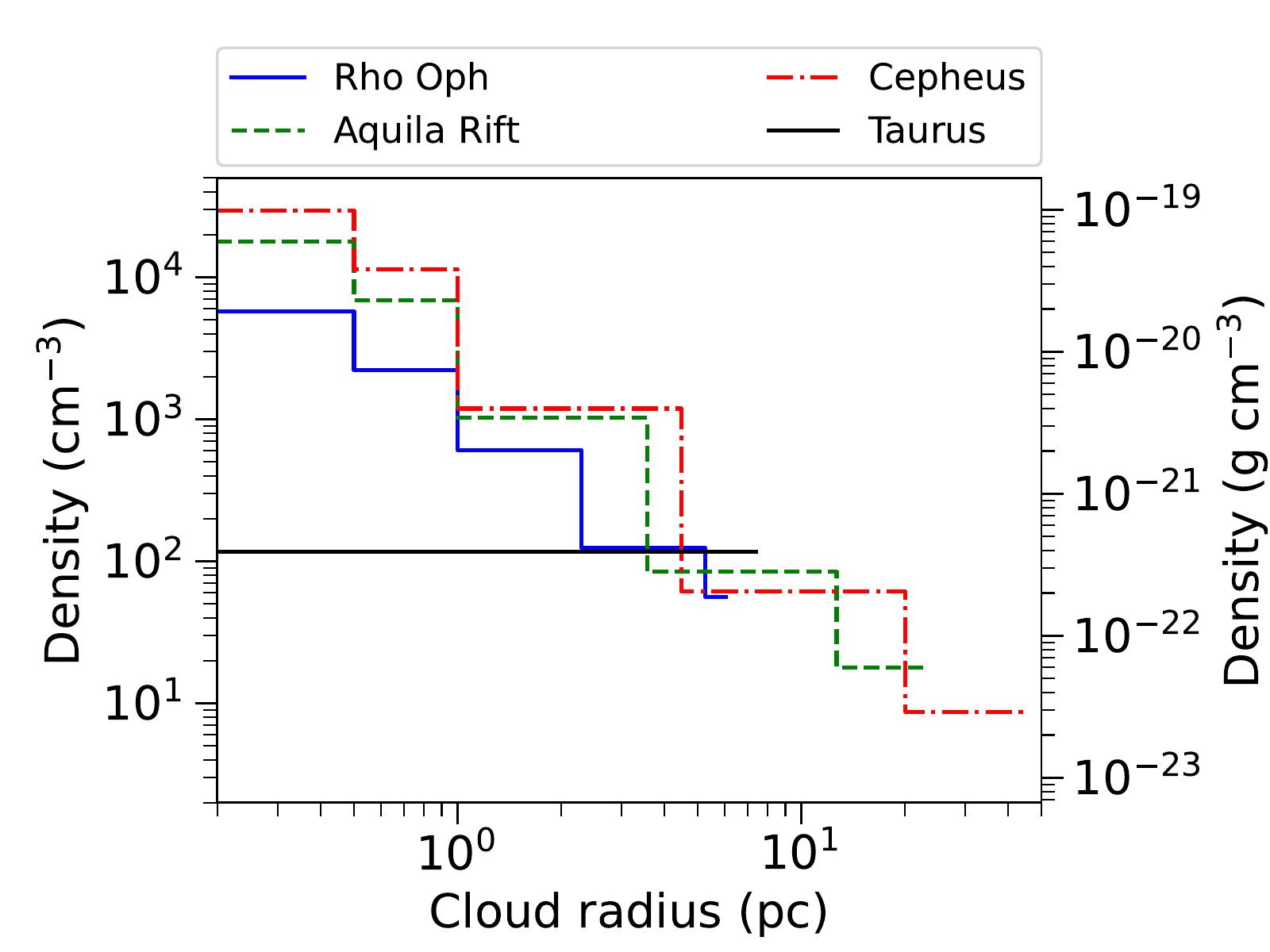}}
\caption{{\it Left panel}: The geometrical model of GMCs consists of five spherical shells and primary GCRs isotropic irradiation is indicated by the black arrows. {\it Right panel}: The gas density variation with cloud radius, from the inner core towards the surface. We have also used a constant gas density for the Taurus cloud to reproduce the gamma-ray flux of  \citep{bagh20} and corresponding plot is shown in the Appendix \ref{sec:repod_T}.}
\label{fig:cloud_dens}
\end{figure}

\subsection{GMC Geometry Model}
\label{GMC_Geometry}
The inherent properties of MCs, such as mass, radius, composition, etc., have a significant impact on the gamma-ray fluxes produced in them. Consequently, we need an accurate model of the cloud. In the simulation, the GMCs were constructed using five spherical concentric shells (Figure \ref{fig:cloud_dens}), each of which is composed of neutral hydrogen ($H_2$) molecules. We assume a gas temperature of $\sim 10$ K, at which most of the gas molecules will be in the neutral state. Inside the GMC, the gas density variation with radius R, i.e., $n_{H_2}(R)$, can be approximated using the relation \citep{gabici2007gamma}

\begin{equation}
    n_{H_2}(R) = \frac{n_0}{1 + \big(\frac{R}{R_0}\big)^{\eta}}~,
    \label{equ:dens}
\end{equation}

where $n_0$ is the density of the molecular hydrogen at the central core of the cloud, and $R_0$ is the radius of the central core. In this simulation, we fixed $R_0$ = 0.5 pc. The number density of hydrogen molecules in each shell of different GMCs is chosen in such a way that the total mass of every GMC reflects the total mass listed in Table \ref{Tab:cloudp} and is different for individual clouds due to differences in their total mass and radius. The variation of gas densities in each GMC shell is shown in Figure \ref{fig:cloud_dens}, where the density is calculated at $R = (R_1 + R_2)/2$, where $R_1$ and $R_2$ are inner and outer radii of the shells. 

The parameter $\eta$ ($0 \leq \eta \leq 2$) basically controls how rapidly the density of the clouds falls \citep{gabici2007gamma}. Therefore, a GMC with a higher value of $\eta$ has a higher value of molecular density in its central region and a lower value of density in its outer region, compared to the same GMC with a lower value of $\eta$. Due to this fact, the value of $\eta$ has a significant effect on the gamma-ray flux coming from the GMC core but can not effectively modify the flux from the whole GMC, as pointed out by \citep{gabici2007gamma}. In this work, the simulations were carried out for two cases with $\eta$ = 1 and $\eta$ = 2, but here we have shown the analyzed results only for the value of $\eta$ = 2 because the obtained gamma-ray fluxes are very similar in the two cases (see Figure \ref{fig:gamma_from_eta} in the Appendix). We don't consider the $\eta = 0$ scenario here to save the computation time and also the constant density case is already discussed by \citep{bagh20}.

\subsection{Physical Interaction Models}
\label{Sec:PhysicalModels}

In GEANT4 a lot of different physical interaction models are available through a set of physics lists which handles the particle interactions and decay channels \citep{geant4ph}. Depending on the types and energy of the projectile particles and target materials, the type of interactions can vary. Typically, the electromagnetic physics lists include the Photoelectric effect, Compton scattering, Gamma conversion, Rayleigh scattering, Coulomb scattering, Pair production, Ionisation, Bremsstrahlung, Annihilation, etc. The hadronic physics lists include different types of Elastic/Inelastic scattering, Capture, and Fission processes. For the current simulation purpose, the $p - p$ inelastic interaction up to very high energy is highly important. This interaction produces $\pi^0$ that eventually decays into two gamma-rays. In this simulation, we used \textit{EmStandardPhysics\_option4}  \citep{geant4phem} from the available physics lists in GEANT4, which can handle electromagnetic interactions up to PeV energies. However, none of the physics lists available in GEANT4 can simulate hadronic interaction above 100 TeV. Consequently, in order to simulate the hadronic interaction at $E>100$ TeV, we used the GEANT4-CRMC interfacing \citep{baus2021cosmic}. The implementation of this interface utilizes the GEANT4 inbuilt hadronic physics list \textit{FTFP\_BERT} \citep{geant4phhdr} up to $\sim 10$ TeV and then switches to \textit{QGSJETII-04} physics list \citep{ostapchenko2011monte} which is valid up to $\sim 10$ PeV.

\begin{figure}[!ht]
    \centering
    \includegraphics[scale=0.65]{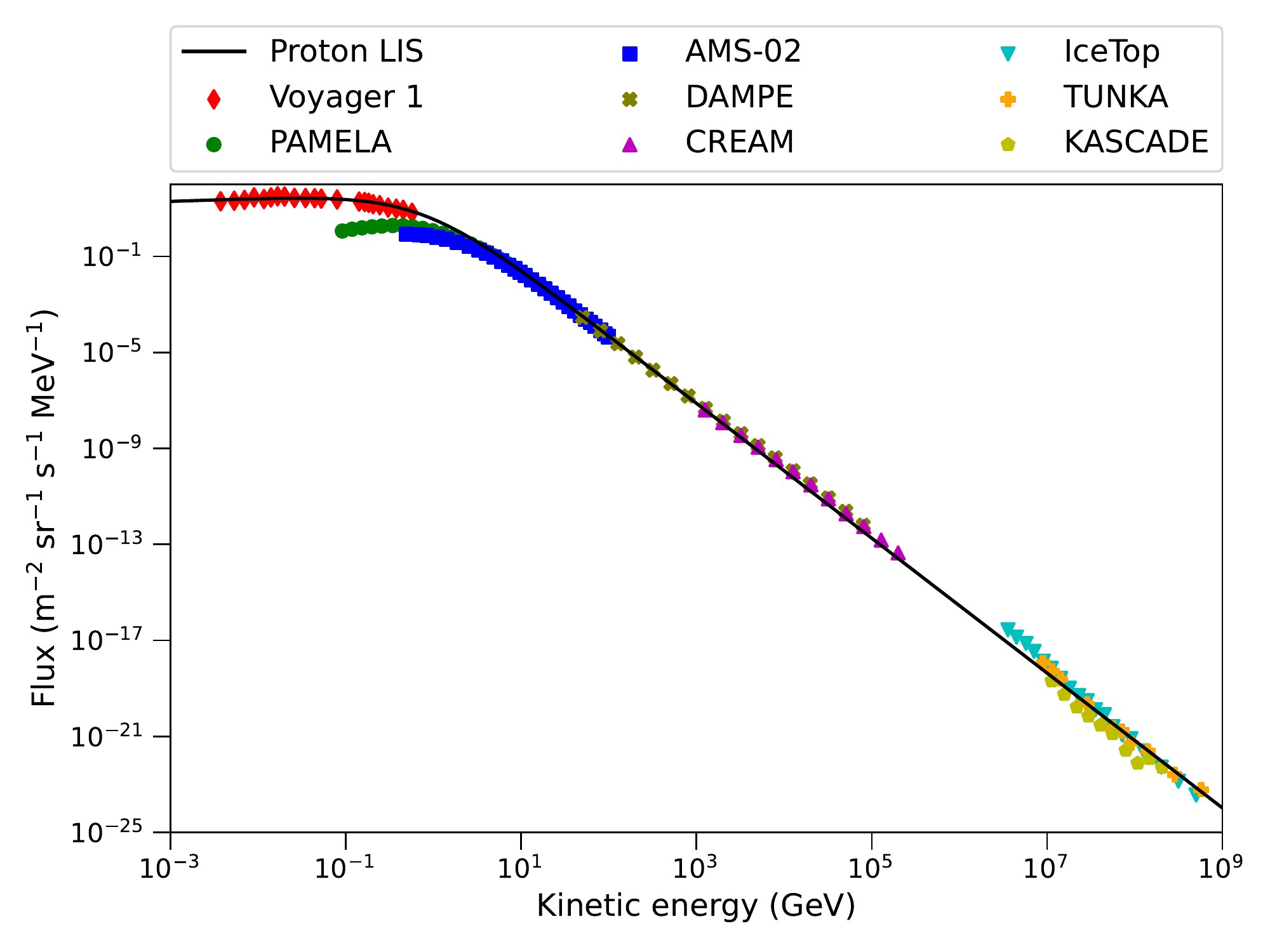}
    \caption{ The diffused spectrum of GCR protons observed on Earth. The data are taken from: Voyager 1 (red points) \citep{stone2013voyager}, PAMELA (green points) \citep{adriani2013time}, AMS-02 (blue points) \citep{aguilar2015precision}, CREAM (magenta points) \citep{yoon2017proton}, IceTop(cyan points) \citep{aartsen2019cosmic}, KASCADE (yellow points) \citep{apel2013kascade}, TUNKA (Orange points) \citep{prosin2014tunka}, DAMPE (olive points) \citep{dampe2019measurement}. The black solid line is a fit to the proton LIS described by Equation \ref{equ:LIS} and based on Voyager 1, PAMELA and AMS-02. The extrapolated curve is extended up to $10^9$ GeV even if we include CR interactions only up to $10^7$ GeV inside GMCs, assuming their Galactic origin.}
    \label{fig:prim_proton}
\end{figure}

\subsection{Primary Particle Generation Model}
\label{Sec:PrimaryPart}

The GCRs are omnipresent in our Galaxy and are considered to be isotropically distributed in local ISM. In 2012, the Voyager 1 spacecraft made the first-ever observation of the distribution of unmodulated CRs in the local ISM \citep{stone2013voyager}. Combined with the observation of AMS-02 \citep{aguilar2015precision} and PAMELA \citep{adriani2013time} space-based CR detectors a new form of primary GCR LIS was proposed by \citep{vos2015new} as shown in Equation \ref{equ:LIS}. 

\begin{equation}
    \Phi_{LIS} = N \frac{E^{1.12}}{\beta^{2}} \Big(\frac{E + 0.67}{1.67} \Big)^{-\alpha},
    \label{equ:LIS}
\end{equation}
where N is the normalization factor (2.7 Particles m$^{-2}$ sr$^{-1}$ s$^{-1}$ MeV$^{-1}$ for proton), E is the kinetic energy (in GeV) of the particle, $\beta$ is particle velocity/speed of light, and $\alpha$ is the spectral index (3.93 for proton).

The distribution of primary GCR protons measured by various space-based detectors is shown in Figure \ref{fig:prim_proton}, along with the calculated LIS. Though the primary GCR is mainly composed of protons, it also contains a small percentage of alpha particles and other heavy elements (HZ). In order to find out the gamma-ray emission from the interaction of alpha and heavy elements we also simulated the interaction of primary GCR alpha particles with the GMCs including an extra contribution of 0.3 fraction added to the primary alpha nucleon flux due to other HZ elements \citep{usoskin2017heliospheric, sarkar2020simulation}. We used the spectral distribution for the alpha particles described in Equation \ref{equ:LIS} --- modifying the parameters --- with $\alpha$ = 3.89, and the normalization factor divided by the nucleon number of alpha particles. The material distributions in the GMCs are then irradiated by the primary particles using the above-mentioned energy distributions in the kinetic energy range of $10^{-1}$ -- $10^7$ GeV from a spherical surface surrounding the GMC geometry. The angular distribution of the particle momenta is done using the cosine-law to ensure the isotropic irradiance of the molecular clouds \citep{zhao2013geometric}. The entire input energy spectrum was divided into four energy intervals ($10^{-1}$--$10^{1}$--$10^{3}$--$10^{5}$--$10^7$ GeV) in order to maintain better simulation statistics at the high energy end where particle flux is lower.  
We simulated $\sim 8 \times 10^8$ incident particles in the whole energy range, i.e., $\sim 2 \times 10^8$ particles in each energy interval. We need to normalize the emitted gamma-ray luminosity ($\Phi_{\gamma}^{S}$) from the simulation with a normalization factor ($N_{F}$) to compare the simulation results to the observed one. Thus the normalized gamma-ray luminosity at the GMC location ($\Phi_{\gamma}^{GMC}$) \citep{banjac2019atmospheric,sullivan1971geometric,roy2021background} is given by

\begin{equation}
    \Phi_{\gamma}^{GMC} = \Phi_{\gamma}^{S} \times N_F = \Phi_{\gamma}^{S} \times \frac{N_R}{N_S},
\end{equation}

where $N_R$ is the integrated values of actual particle flux at the source location and $N_{S}$ is the number of the simulated events in the energy range under consideration (e.g., $E_1 = 10^{-1}$ GeV and $E_2 = 10^7$ GeV). $N_R$ is calculated by integrating the incident primary GCR LIS over the considered energy range, the area of the surface generating the particles, and the solid angle of the angular distribution.

\begin{equation}
    N_R = \int_{E_1}^{E_2} \Phi_{LIS}\;dE \int_{0}^{\pi/2} {\rm cos}\theta\; {\rm sin} \theta\;d\theta \int_{0}^{2\pi} d\phi \; \int_{s}\;ds,
\end{equation}

Now, to compare the normalized gamma-ray luminosity ($\Phi_{\gamma}^{GMC}$) --- coming out of the GMCs --- with the gamma-ray flux measured by the Fermi-LAT satellite, we must divide $\Phi_{\gamma}^{GMC}$ by $4\pi D^2$ (i.e., total surface area subtended at the position of the observer, where D is the distance of the GMC from the observer). So, the observed gamma-ray flux near the Earth ($\Phi_{\gamma}^{O}$) is given by

\begin{equation}
    \Phi_{\gamma}^{O} = \frac{\Phi_{\gamma}^{\rm GMC}}{4\pi D^2 },
\end{equation}


\section{Results and Discussion}
\label{Sec:Discussion}

\begin{figure}[!ht]
\centering
\includegraphics[width=\linewidth]{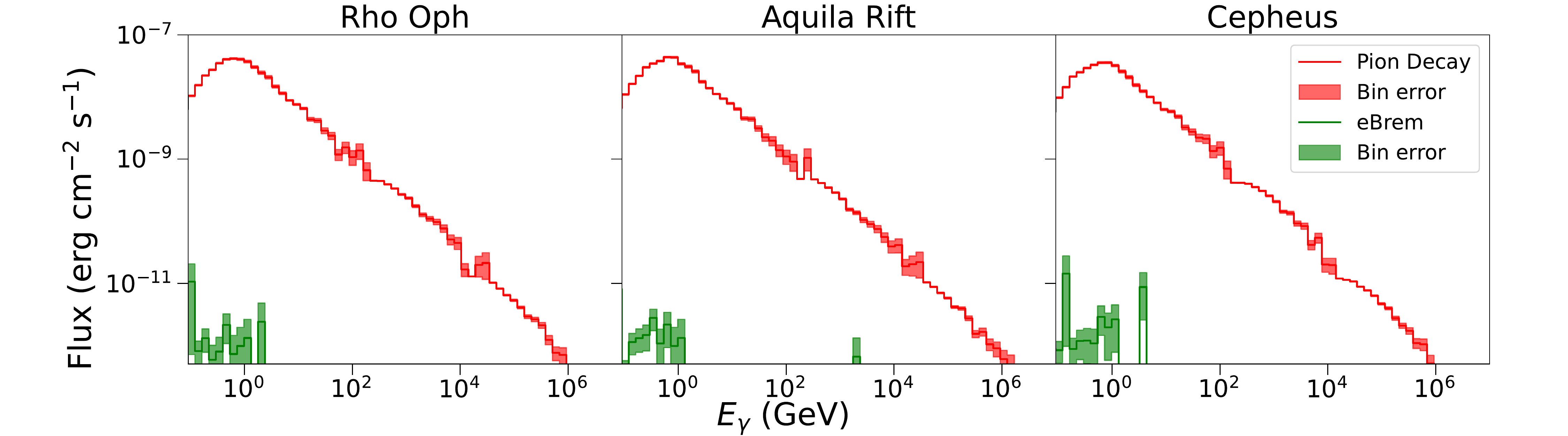}
\caption{The simulated gamma-rays originated from the three different MCs (Rho Oph, Aquila Rift, and Cepheus) at the cloud location. The gamma-ray flux here is equivalent to luminosity $\Phi^{GMC}_{\gamma}$ per unit area of the GMCs}. The red curve shows the gamma-ray flux from $\pi^0$ decay and the green curve is for bremsstrahlung radiation due to electrons, produced via $p-p$ interactions. $\pi^0$ decay is the dominant channel for the production of gamma-ray flux inside GMCs.
\label{fig:gamma_process}
\end{figure}

The total gamma-ray flux produced from the interaction of primary GCR protons with each of the individual GMCs is displayed in Figure \ref{fig:gamma_process}. The figure shows that a primary GCR proton with an energy of 10 PeV can readily create a PeV gamma-ray from GMCs, where the Spectral Energy Distributions (SED) for each GMC are quite similar both in terms of spectral shape and intensity. This could be due to the fact that the GMCs are irradiated with the same particle flux and comparable densities between their outer shells, with the exception of Aquila Rift and Cepheus which have relatively few molecules per cubic centimetre. Because of fewer target molecules available for interactions in their outer shells, the contribution of these to the total gamma-ray flux is small relative to the high-density shells. The larger error bars at around 10$^2$, 10$^4$, and 10$^6$ GeV energy come from our choice to divide the whole simulated energy range into four sub-energy ranges. The error bars are not exactly at the incident energy divisions (i.e., $10^{-1}$--$10^{1}$--$10^{3}$--$10^{5}$--$10^7$ GeV) because the secondary photons get $\sim$ 10\% of the kinetic energy of primary protons in $\pi^0$ decay process.

From our simulation study, we have also shown that almost all of the gamma-rays created in our energy range of interest (100 MeV to 10 PeV) are through the \textit{$\pi^0$ decay} process and that the contribution of \textit{Bremsstrahlung} emission from the electrons generated by $p-p$ interactions are insignificant. This is shown in Figure \ref{fig:gamma_process}.

\begin{figure}[!ht]
    \centering
    \includegraphics[scale=0.65]{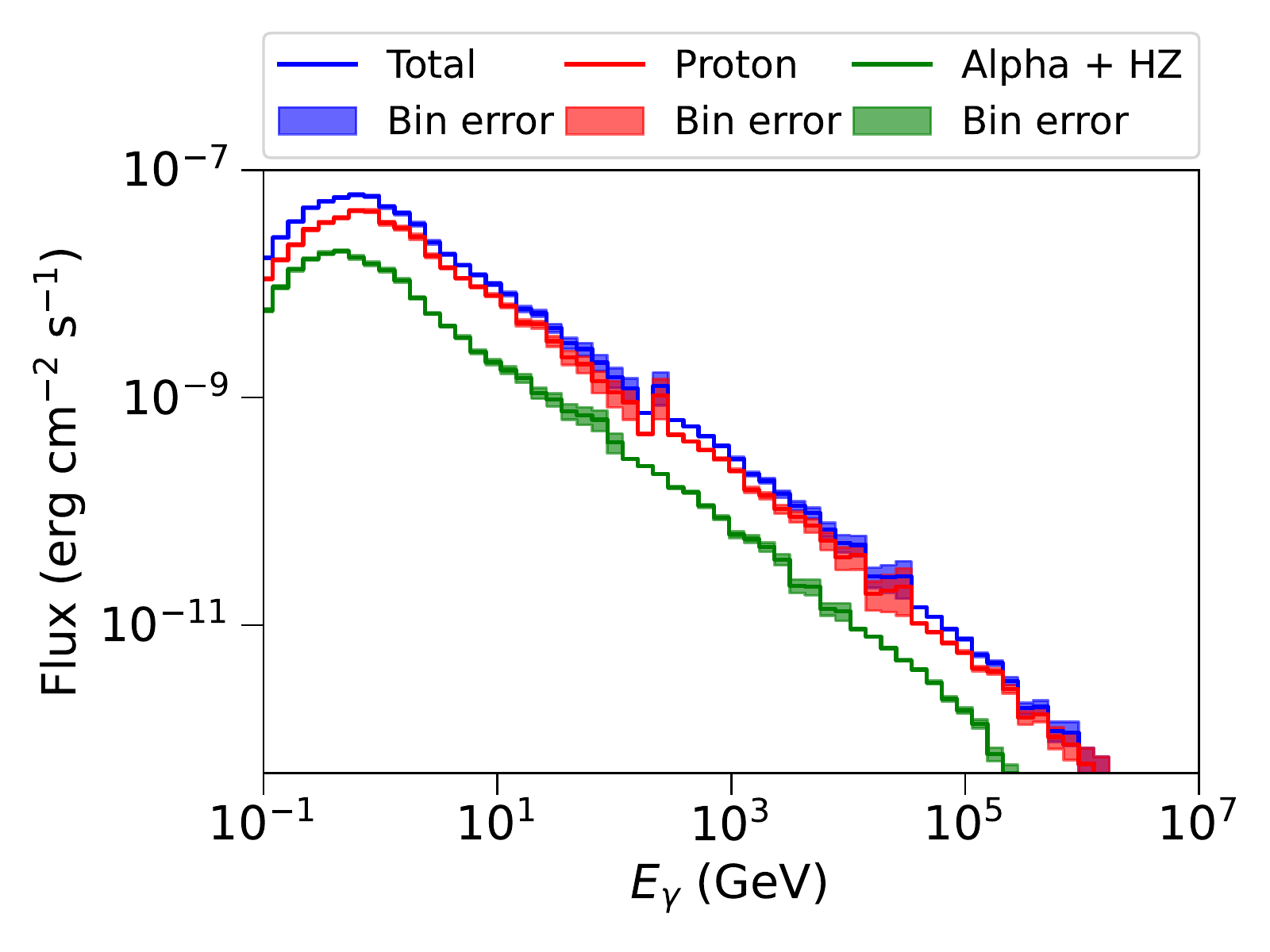}
    \caption{ Production of total gamma-rays from the primary GCR protons and heavy nuclei with the gas density available in the Aquila Rift GMC.}
\label{fig:gamma_from_pna}
\end{figure}

The contributions from the HZ elements in the total observed gamma-ray flux has already been studied by several authors \citep{huang2007gamma, kachelriess2014nuclear, kafexhiu2014parametrization, mori1997galactic, mori2009nuclear}. They found that a Nuclear Enhance Factor (NEF) in the range of 1.45 to 2.1 (above GeVs) needs to be multiplied by the gamma-ray flux produced from the primary GCR proton to find the total gamma-ray emission due to all CR nuclei. It is generally used as an energy-independent constant factor, but it can vary depending on the energy of incident particles. Using this simulation framework, we have also tried to find out the NEF for heavy elements in hydrogen molecular targets. In Figure \ref{fig:gamma_from_pna}, the total gamma-ray flux due to CR protons and also due to heavy nuclei is shown for Aquila Rift GMC. \cite{mori2009nuclear} assumed hydrogen as the only target element and found the value of NEF $\sim 1.3$ at CR energy 10 GeV/nucleon. Based on our simulation, we have found the average NEF from the heavy elements in the energy range of Fermi-LAT observation (i.e., 3--1000 GeV) is $\sim$ 1.32. However, it should be noted that heavy elements present in the ISM also contribute to the total gamma-ray flux. But this has not been considered in this simulation. So, in order to include the effect of this contribution too, we consider a total NEF of 2.09, as suggested by \cite{kachelriess2014nuclear}.

\begin{figure}[!ht]
\centering
\includegraphics[width=\linewidth]{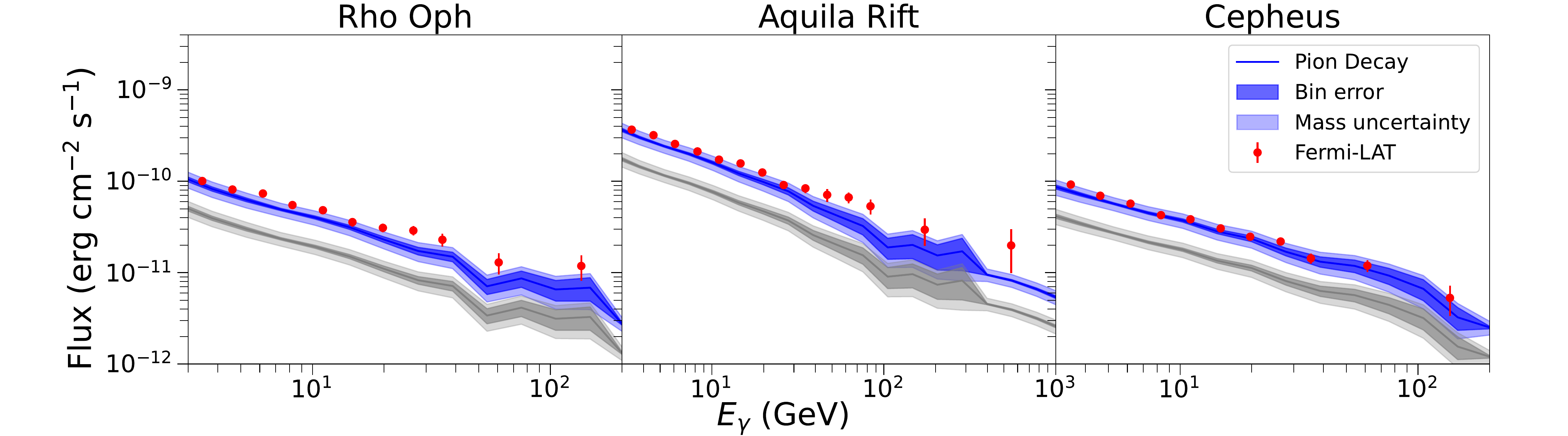}
\caption{Comparison between simulated gamma-ray spectra originated from the interaction of primary GCRs with the three GMCs (Rho Oph, Aquila Rift, and Cepheus) and the observed gamma-ray spectra from these regions by Fermi-LAT satellite \citep{bagh20}. The grey curve here shows the simulated gamma-rays resulting from the interaction of primary GCR protons with the molecular hydrogen present in the GMCs and the blue curve represents re-scaled simulated gamma-ray flux by a factor of 2.09 to take into account the contributions from the heavy elements present in both the ISM and GCRs.}
\label{fig:gamma_comp_flat}
\end{figure}

In Figure \ref{fig:gamma_comp_flat}, we show the simulated gamma-ray flux inside the three GMCs (Rho Oph, Aquila Rift, and Cepheus) due to GCRs compared with the Fermi-LAT observations. The grey curve represents the gamma-ray flux due to the interaction of primary GCR protons with the hydrogen molecules inside the GMCs. The blue curve represents the same flux multiplied with a constant NEF of 2.09 --- including the nuclei contribution.

\begin{figure}[!ht]
\centering
\includegraphics[width=0.65\linewidth]{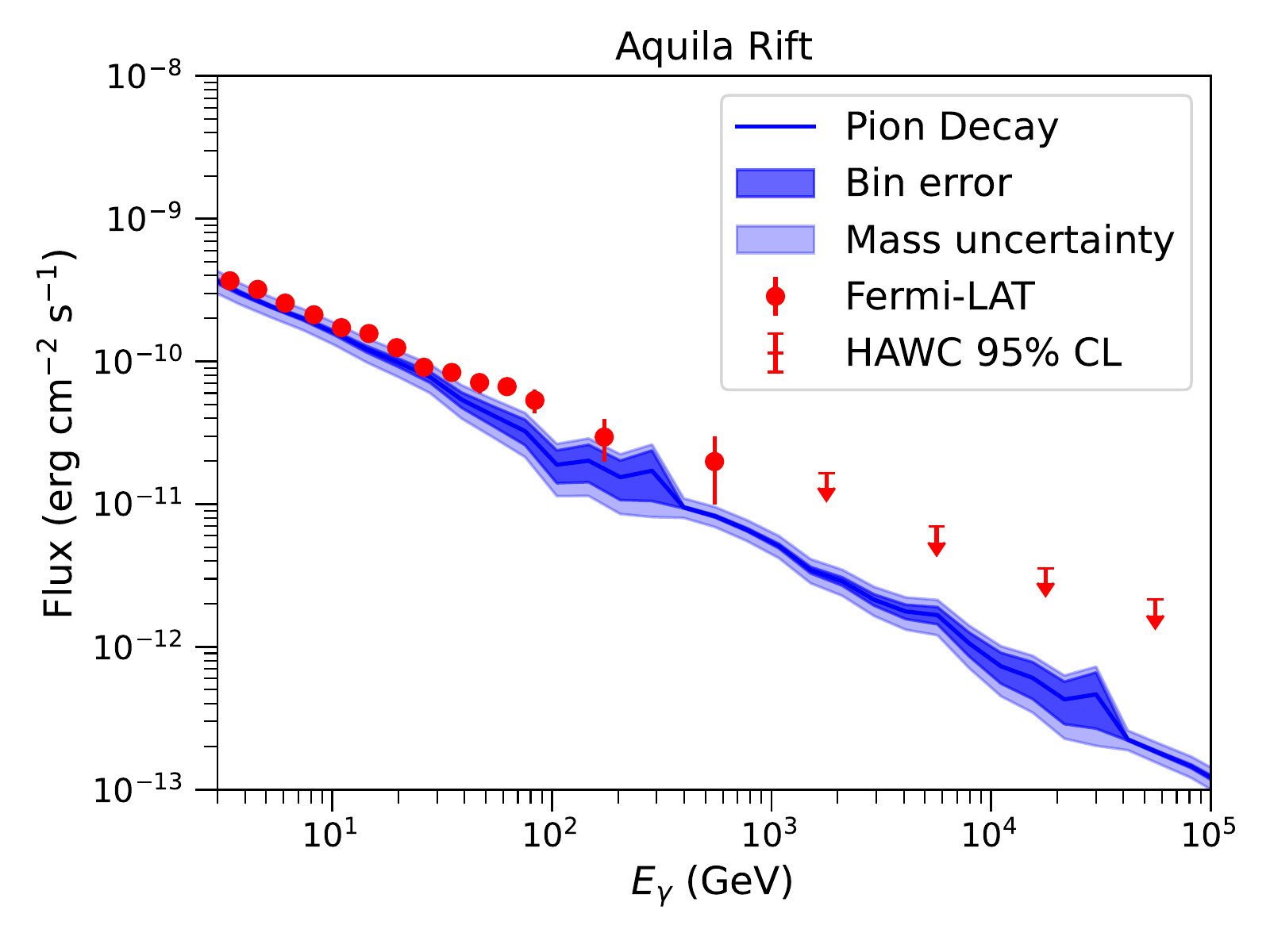}
\caption{Comparison of the simulated gamma-rays from the Aquila Rift in GeV-PeV range. The GeV-TeV data is from the Fermi-LAT satellite \citep{bagh20} and the upper limits (95\% confidence level upper limits) estimated by HAWC observation \citep{albert2021probing}.}
\label{fig:gamma_comp_hawc}
\end{figure}

For the Aquila Rift, HAWC upper limits are available for gamma-rays in the TeV-PeV range \citep{albert2021probing}. For comparison purposes, we calculated the gamma-ray flux from the Aquila Rift also in this energy range. The overlaying plot of the simulated gamma-rays with the observed results from the Aquila Rift GMC region is shown in Figure \ref{fig:gamma_comp_hawc}.

\begin{figure}[!ht]
\centering
\includegraphics[width=0.65\linewidth]{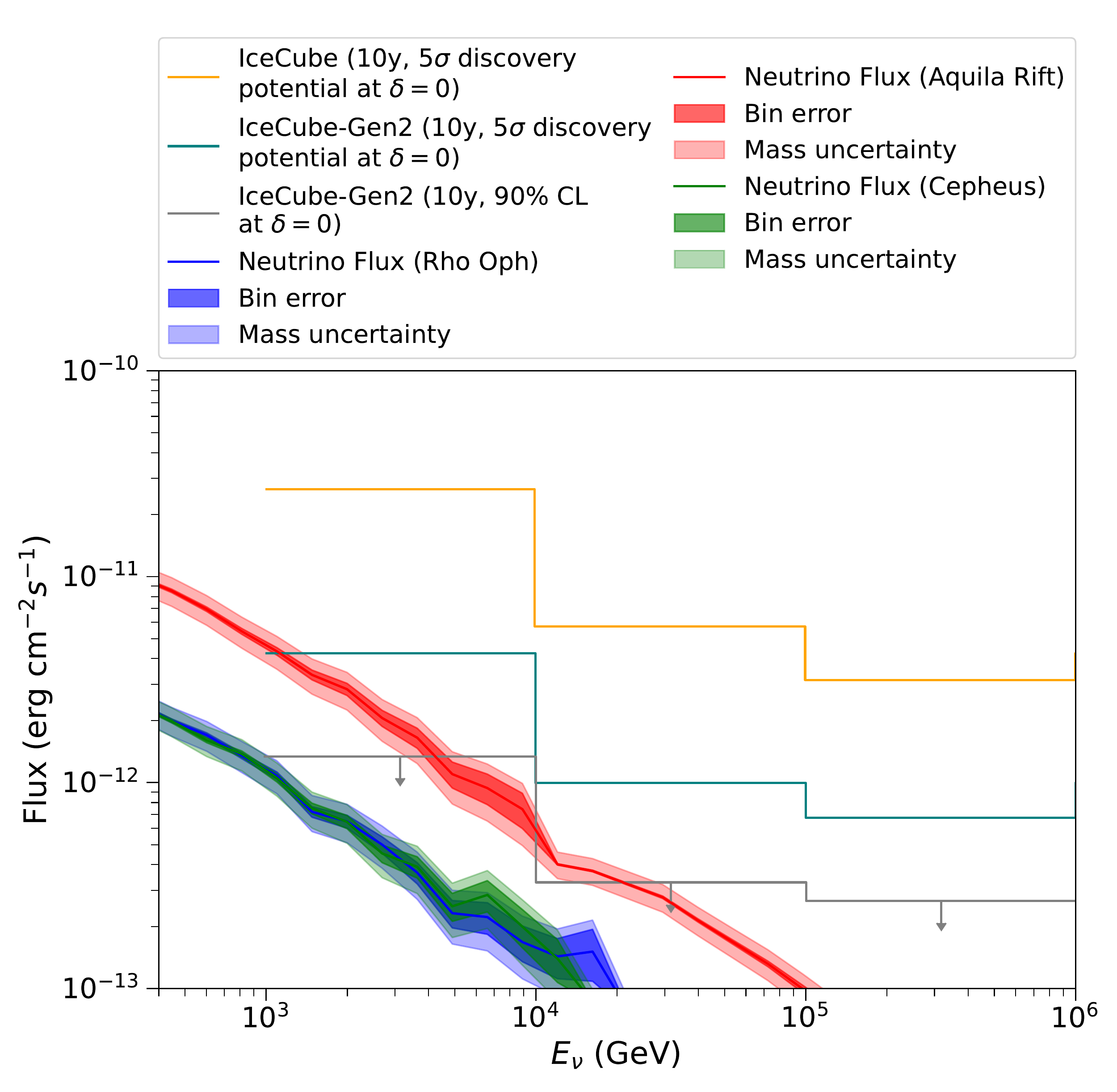}
\caption{The total simulated neutrino ($\nu_{e,\mu}$ + $\bar{\nu}_{e,\mu}$) flux from the Rho Oph, Aquila Rift, and Cepheus MCs. The IceCube sensitivity is also shown for its first and second-generation (orange and teal line) telescopes. \citep{aartsen2021icecube}.}
\label{fig:neutrino_iceCube}
\end{figure}

The production of gamma-rays from interactions of GCRs and GMCs through a hadronic $p-p$ process is accompanied by the emission of neutrinos. These neutrinos are mainly produced from the decay of charged pions and muons. For this reason, we expect the flavour ratio at the position of the source to be $\nu_e : \nu_{\mu} : \nu_{\tau} = 1 : 2 : 0$. However, due to the neutrino oscillation, the flavour ratio at the observation location here on the Earth will be modified into $\nu_e : \nu_{\mu} : \nu_{\tau} \approx 1 : 1 : 1$ \citep{aartsen2015combined}. In Figure \ref{fig:neutrino_iceCube}, the sensitivity of the IceCube detector for 5$\sigma$ discovery potential at the celestial equator over ten years of observation is shown (orange line) along with the total simulated neutrino fluxes from the three GMCs. The total neutrino fluxes shown here include both the contributions from $\nu_{e,\mu}$ + $\bar{\nu}_{e,\mu}$, but we do not find any contributions from $\nu_{\tau}$ + $\bar{\nu}_{\tau}$. It is also evident from the figure that the current generation of IceCube detectors is not sensitive enough to detect neutrino signals from GMCs. However, the sensitivity of the proposed future-generation IceCube-Gen2 detectors (teal line) will be sufficient enough to detect the signals from some of the large and nearby GMCs.


\section{Summary and Conclusions}
\label{Conclusions}

We have discussed the total gamma-ray and neutrino flux produced in three particular Galactic GMCs through the interaction of GCRs. The gamma-ray spectrum estimated in this work can explain the Fermi-LAT data for all GMCs quite satisfactorily. However, for the Rho Oph and Aquila Rift GMCs the spectra are harder in between 30--100 GeV, and its origin might be due to primary Proton/Helium particle spectra above $\sim$ 200 GeV \citep{adriani2011pamela, aguilar2015precisionalpha}, which has also been reported by \citep{bagh20}. Nevertheless, our model can explain the gamma-ray spectra of Cepheus GMC quite well.  Another probable origin of this hardening in the gamma-ray spectra of GMCs could be  due to CR re-acceleration in the vicinity of GMCs \citep{2016AnA...595A..58C}. In the future, we can study the impact of these physical scenarios by including them in the current simulation framework.

Future observations with the next-generation Imaging Atmospheric Cherenkov Telescopes, as ASTRI Mini-Array \citep{pareschi2013dual,scuderi2019astri,giro2019astri,scuderi2022astri,vercellone2022astri} or Cherenkov Telescope Array (CTA) \citep{cherenkov2019science,acero2023sensitivity} would be important to reveal any point source populations that can contribute to gamma-rays above a few GeV from these GMCs. The estimated neutrino flux in this work for the GMCs under consideration has been compared with the current IceCube sensitivity in Figure \ref{fig:neutrino_iceCube}. The Aquila Rift source looks like a potential candidate for the second-generation IceCube neutrino observatory, due to its larger mass and its relatively short distance (hence larger $M/D^2$ value) as given in Table \ref{Tab:cloudp}. Its neutrino flux is comparable to the sensitivity of the second-generation IceCube neutrino observatory in the range of 1--10 TeV.

\section*{Acknowledgements}
 The authors are thankful to the anonymous referee for the insightful comments on our work that helped us to improve the manuscript. JCJ is thankful to S. Chakraborty for helpful discussions. JCJ was partially supported by a grant from the University of Johannesburg Research Council.

\bibliography{ref_prop}
\bibliographystyle{JHEP}

\appendix

\counterwithin{figure}{section}
\section{APPENDIX: Effect of $\eta$ on the Secondary Gamma-ray Flux}

\begin{figure}[!ht]
\centering
\subfigure{\includegraphics[width=0.45\linewidth]{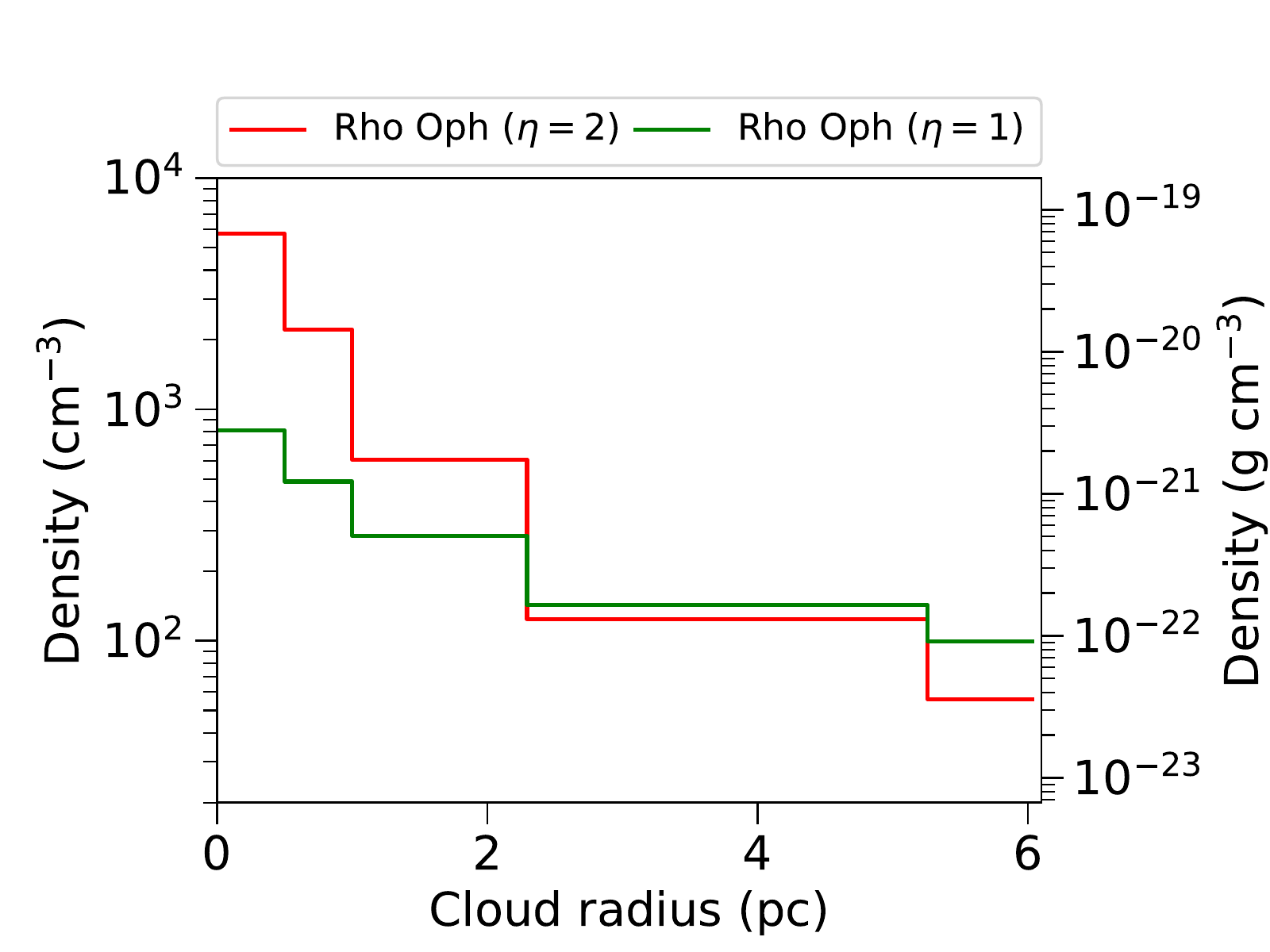}}
\subfigure{\includegraphics[width=0.45\linewidth]{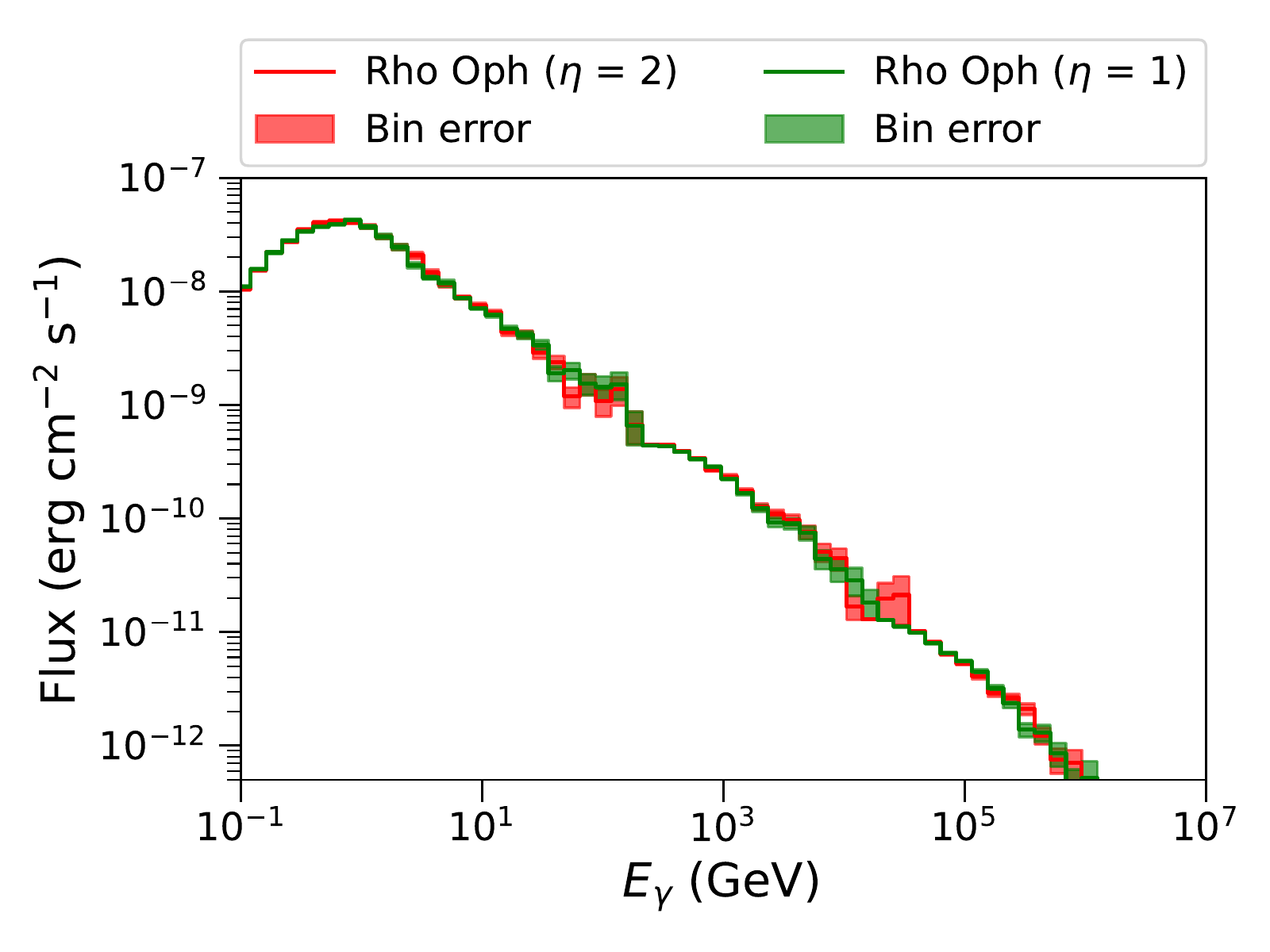}}
\caption{{\it (a) Left Panel}: Variation of the molecular density in Rho Oph GMC with $\eta$ = 2 and 1. {\it (b) Right Panel}: $\pi^{0}$ decay gamma-ray flux from Rho Oph using the density distributions as shown in the left panel: no appreciable changes are observed. }
\label{fig:gamma_from_eta}
\end{figure}

We have simulated the interaction of the primary GCR protons with Rho Oph GMC for two different values of $\eta$ in Equation \ref{equ:dens} to find out how the variation in density of each shell of a GMC could modify the production of gamma-rays from the whole GMC. The simulated gamma-ray fluxes, as shown in Figure \ref{fig:gamma_from_eta}b, however, do not show any significant difference for the two values of $\eta$. This is also mentioned in \citep{gabici2007gamma}, that the change in the density profile could not significantly modify the emission of gamma-rays from the whole cloud. This could be due to similar densities in the outer shells of the GMCs (see Figure \ref{fig:gamma_from_eta}a).

\section{APPENDIX: Gamma-Ray Flux from Taurus GMC with Constant Gas Density Profile}
\label{sec:repod_T}

\begin{figure}[!ht]
\includegraphics[scale=0.65]{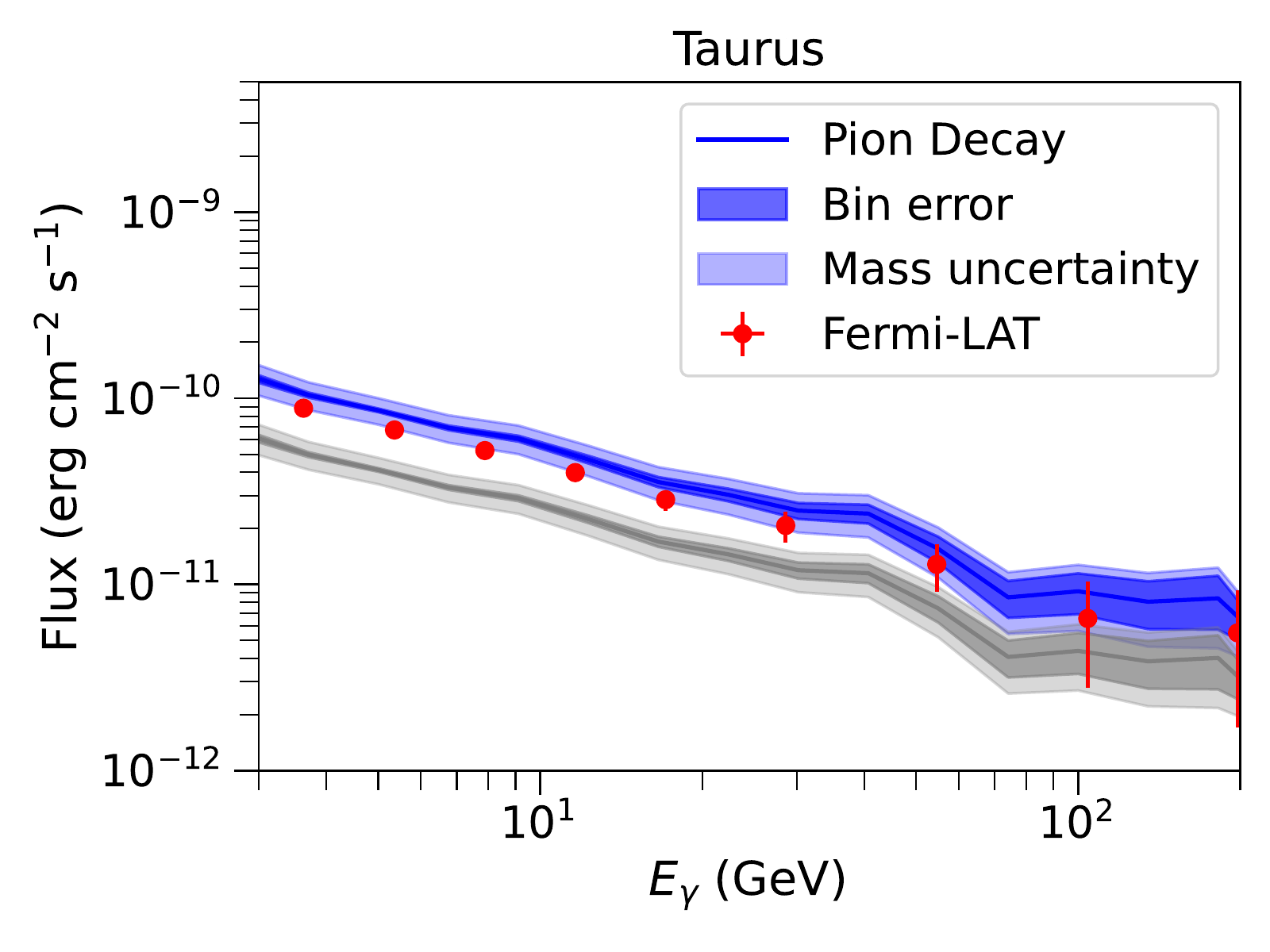}
\caption{This plot shows the reproduced gamma-ray flux for Taurus GMC as estimated by \citep{bagh20}, and the consistent calculation by GEANT4 simulation setup.}
\label{fig:gamma_taurus_rep}
\end{figure}

\end{document}